\documentclass[12pt,preprint]{aastex}
\usepackage{natbib}
\def\msun{\mbox{$M_\odot$}}

\shorttitle{CHANDRA OBSERVATION OF ABELL 2063}
\shortauthors{KANOV, SARAZIN, \& HICKS}
\slugcomment{Astrophysical Journal, in press}

\begin{document}

\title{Chandra Observation of the Interaction of the Radio Source
and Cooling Core in Abell 2063}

\author{Kalin N. Kanov,
Craig L. Sarazin,
and
Amalia K. Hicks}

\affil{Department of Astronomy, University of Virginia, \\
P. O. Box 400325, Charlottesville, VA 22904-4325; \\
knk5n@virginia.edu,
sarazin@virginia.edu,
ahicks@alum.mit.edu}

\begin{abstract}
We present the results of a {\it Chandra} observation of the cooling core cluster
Abell 2063.
Spectral analysis shows that there is cool gas (2 keV) associated with
the cluster core, which is more than a factor of 2 cooler than the outer
cluster gas (4.1 keV).
There also is spectral evidence for a weak cooling flow,
$\dot M \approx 20 \, \msun$ yr$^{-1}$.
The cluster exhibits a complex structure in the center that consists of
several bright knots of emission,
a depression in the emission to the north of the center of the cluster,
and a shell of emission surrounding it.
The depression in the X-ray emission is coincident with
the position of the north-eastern radio lobe of the radio source associated
with the cluster-central galaxy.
The shell surrounding this region appears to be hotter,
which may be the result of a shock that has been driven into the gas by
the radio source.
The power output of the radio source appears to be sufficient to offset
the cooling flow,
and heating of the gas through shocks is a possible explanation
of how the energy transfer is established.
\end{abstract}


\section{Introduction} \label{sec:intro}

Clusters of galaxies contain very hot gas, the intracluster medium (ICM).
Due to the extremely strong gravitational fields of clusters, plasma that falls
towards the center will gain significant gravitational potential energy, which
results in temperatures in the ICM of $\sim$$10^{8}$ K.
With time, energy is lost due to X-ray radiation (thermal bremsstrahlung
and line emission from metals) from the ICM,
which acts to cool the X-ray emitting gas unless these energy losses are
balanced by some heating process.
One would expect this cooling gas to flow toward the dense core of the
cluster.
The cooling process is known as a ``cooling flow.''
This process can potentially trigger star formation in the core.
Evidence for such cooling flows have been
observed by the {\it Einstein}, {\it ROSAT}, and {\it ASCA} observatories
\citep[e.g.,][]{fabian94}.
The predicted cooling rates from these observations are hundreds of solar
masses per year.
More recent observations made with the {\it XMM-Newton} and {\it Chandra} 
X-ray observatories suggest cooling rates that are at least an order of 
magnitude lower \citep[e.g.,][]{peterson03}, with most of the gas only 
cooling down to about one third of the ambient cluster temperature.

One of the proposed explanations of this phenomenon and a solution to the
discrepancies in measured cooling rates is heating of the ICM by a central
radio source.
These radio sources are thought to be the result of accretion
events by active galactic nuclei (AGN) within the central cD galaxies
that are generally present in the centers of cooling core clusters.
X-ray observations show that the lobes of the central radio sources
can displace the X-ray emitting gas and leave ``holes'' in the ICM, which
are filled by the synchrotron plasma
\citep[e.g.,][]{fabian00}.
Some of the theoretical models for these ``radio bubbles'' predicted that
the ICM would be heated by strong shocks from the radio emitting plasma
\citep[e.g.,][]{heinz98}.
However, in most cases the observed shells of
higher surface brightness surrounding the holes have been found to have lower
temperatures than the surrounding gas
\citep[e.g.,][]{blanton01}.
More recently, a few cases of radio lobes bounded by strong shocks have
been observed
\citep{mcnamara05,nulsen05a,nulsen05b}.
In addition, weak shocks or sound waves have been detected around the radio
lobes in Perseus
\citep{fabian03}.
The energy output contained in the radio outbursts is
found to be sufficient to counterbalance the cooling flow in many cases
\citep[e.g.,][]{birzan04}.

In this paper, we present an analysis of a {\it Chandra} observation of the
cooling core cluster Abell~2063.
Abell~2063 is a rich cluster of galaxies at a redshift of $z = 0.0349$.
The central cD galaxy in the cluster is host to the radio source
[OL97]1520+087 \citep{owen}.
Abell~2063 has been previously observed in the X-ray with
{\it Einstein} \citep{david93,jf84},
{\it ROSAT} \citep{david95,peres98}, and
{\it ASCA} \citep{white00}.
The derived temperature from the {\it Einstein} and {\it ROSAT} data
is $kT = 4.1$ keV,
whereas the {\it ASCA} data gives a temperature of $kT = 3.9$ keV and
an average chemical abundance of 0.24 times the solar value.
The bolometric luminosity as computed from the {\it Einstein} data is
$L_{\rm bol} = 1.48 \times 10^{44}$ ergs s$^{-1}$.
The cooling rate derived from
these observations is $\dot M=50^{+39}_{-47} \, M_{\odot}\, {\rm yr}^{-1}$
\citep{white00} for the {\it ASCA} data and
$\dot M = 19^{+4}_{-6} \, M_{\odot}\, {\rm yr}^{-1}$ \citep{peres98}
within a radius of $r_{\rm cool} = 68^{+9}_{-21}$ kpc for the {\it ROSAT} data.
Throughout this paper we adopt WMAP cosmological
parameters \citep{wmap}
of $H_0$ = 71 km s$^{-1}$ Mpc$^{-1}$,
$\Omega_{\Lambda}=0.73$, and $\Omega_M=0.27$. At the redshift of
Abell~2063, this corresponds to a scale of 0.686 kpc/arcsec. The
uncertainties quoted in this paper are 90\% confidence intervals unless
otherwise stated.

\section{Observation and Data Reduction} \label{sec:data}

Abell~2063 was observed with the {\it Chandra} ACIS-S3 detector.
The initial observation was interrupted due to a strong solar flare, and,
as a result, the data were taken in three intervals of
14,162 s on 2005 March 30, 16,824 s on 2005 April 1, and 9906 s on 2005
April 5. The nominal values of the pointing direction and roll angle were kept
the same for all three observations, so that the instrument field-of-view (FOV)
is very nearly identical for the three observations. The pointing was chosen so
that the cluster core was in the FOV of the ACIS-S3 detector, and was 1\arcmin\
away from the center of the detector in order to avoid the node boundaries
on the chip.
The event mode of the observation was a timed exposure in very faint (VF)
mode.
The temperature of the CCD was $-120$ C and the frame times were 3.14 s.
Only events with grades of 0, 2, 3, 4, and 6 were accepted.
In addition, the VF mode grades were used to reject additional particle
background\footnote{http://cxc.harvard.edu/cal/Acis/Cal\_prods/vfbkgrnd/}.
Data was collected from chips 3, 5, 6, 7 and 8, but we only included
data from chip 7 (ACIS-S3) in the analysis of the cluster.
We used data from chip 5 (ACIS-S1) to check for background flares.

The data reduction was done with the help of the software package CIAO,
version 3.2.1\footnote{http://cxc.harvard.edu/ciao/}.
An observation specific bad pixel file was created using the tool
acis\_run\_hotpix and applied to the data.
The count rates in the energy band 2.5--6.0 keV on the ACIS-S1 chip were used
to check for background flares, using the
lc\_clean script\footnote{http://cxc.harvard.edu/contrib/maxim/bg/}.
The data was clipped on both sides of the mean if
there was a difference of more than a factor of 1.2 from the mean.
This resulted in excluding 5094 s of data from the first observation,
1001 s of data from the second observation,
and 17 s of data from the third observation,
leaving a cleaned total exposure of 34,779 s.
However, subsequent analysis indicated that the background was elevated,
particularly at moderate event energies, for all of the first observation.
This didn't significantly affect the images of the center of the cluster
where the surface brightness is very high.
However, it would have been difficult to model the spectrum of this extra
background component. Thus, all of the spectral analysis was done excluding the
first observation.
The total exposure for spectral analysis was 25,711 s.

Since the pointings of the three observations were essentially identical,
the events files were merged into a single file using the CIAO script
merge\_all. Before doing this, the sources in the FOV of the S3 in each of the
individual observations were detected with the CIAO tool wavdetect.
The source positions were compared to check for any small systematic offsets.
The positions of sources in the first two observations agreed very
accurately, but they were shifted very slightly in the third observation.
The positions in the third observation were shifted by $+0\farcs16$ in RA and
$-0\farcs09$ in Dec.

A background events file was created from the blank-sky observations of M.
Markevitch, which are also included in the CIAO calibration database (CALDB).
A stack of the aspect solutions for the observations was used and the events
from the blank-sky observations were reprocessed using the CIAO tool
reproject\_events to match the observation at hand.
The background exposure was
normalized by the ratio of the blank-sky background rate over the observation
background rate for PHA channels 2500--3000 (roughly, photon energies of
9--12 keV).

\section{X-ray Image} \label{sec:image}

\begin{figure}[hp]
\vskip 4.0truein
\includegraphics{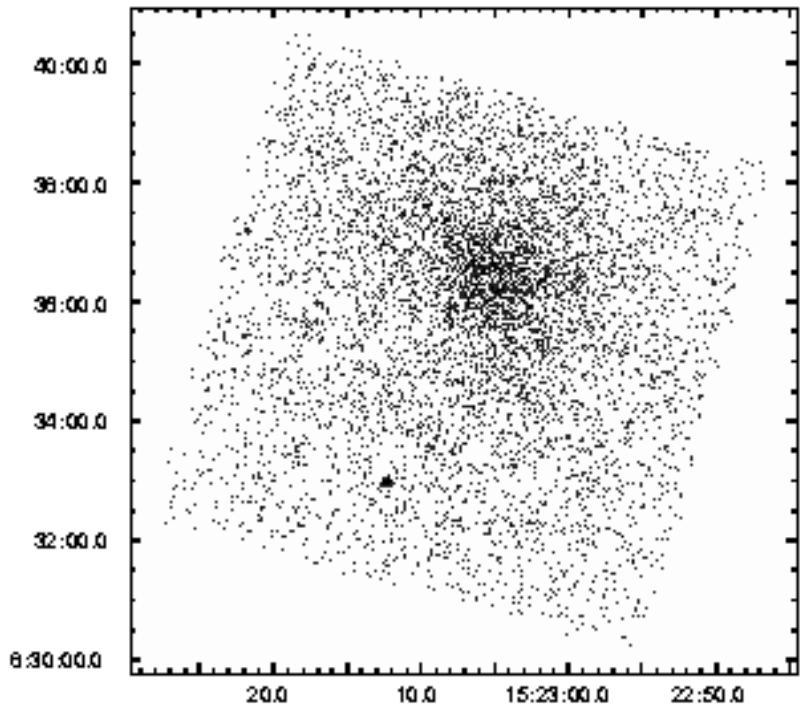}
\vskip 0.1truein
\figcaption{Raw $Chandra$ image of the full ACIS-S3 chip ($\sim 346
\times 346$ kpc) region of Abell 2063 in the $0.3-10.0$ keV energy
band. The image has not been corrected for background or exposure.
The X-ray emission is approximately circular in shape and
there is evidence for structure in the center.
The three stripes of fainter emission are node boundaries on the S3 chip;
the central one is also augmented by several bad columns.
The coordinates are J2000.
\label{image:raw_image}}
\end{figure}

\begin{figure}[hp]
\vskip 4.0truein
\includegraphics{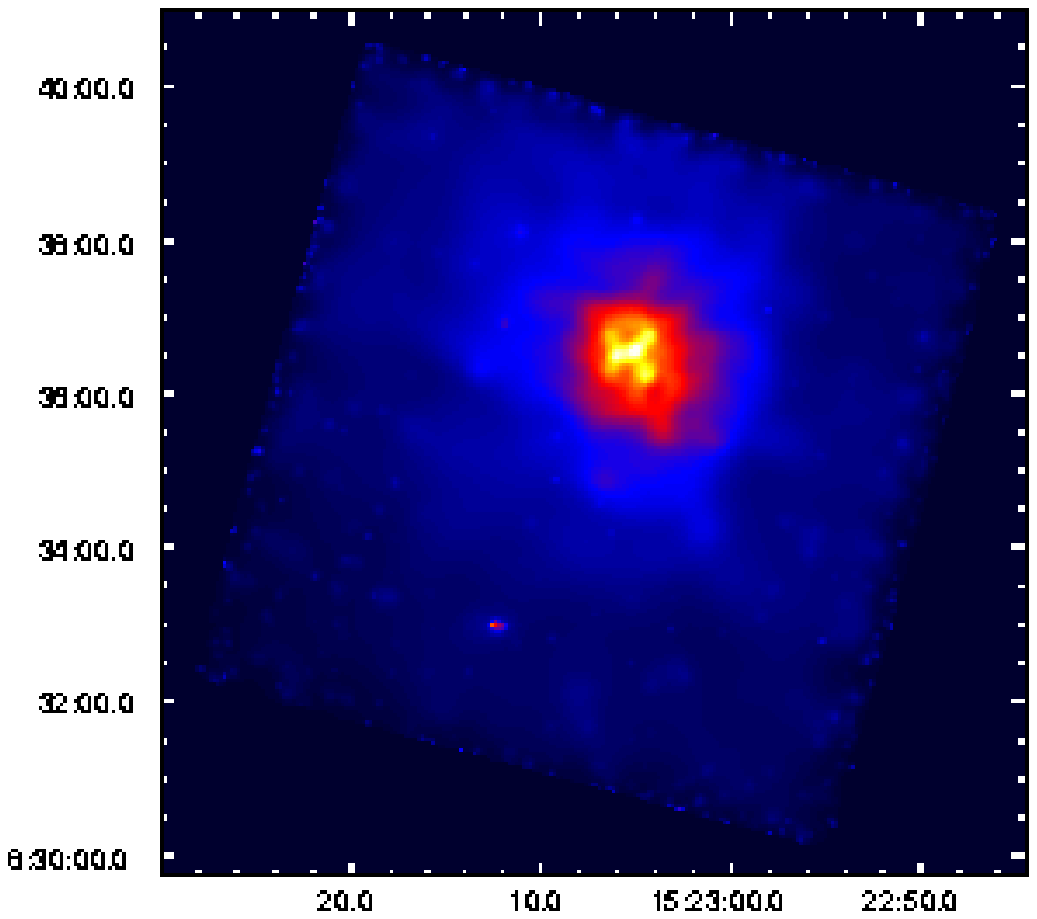}
\vskip 0.1truein
\figcaption{Adaptively smoothed image of the full ACIS-S3 chip ($\sim 346
\times 346$ kpc) region of Abell 2063 in the $0.3-10.0$ keV energy
band.
The image was smoothed to a minimum signal-to-noise of 3 per smoothing
beam, and corrected for background and exposure. The surface 
brightness color scale ranges approximately from $2.7 \times 10^{-7}$ (dark blue) 
to $3.9 \times 10^{-4}$ (yellow-white) counts s$^{-1}$ pixel$^{-1}$. 
\label{image:smoothed_image_S3}}
\end{figure}

Figure~\ref{image:raw_image} shows the raw X-ray image from the {\it Chandra}
S3 chip in the 0.3--10.0 keV band.
It has not been corrected for background or exposure and has not been smoothed.
The central bright region is approximately circular in shape.
Figure~\ref{image:smoothed_image_S3} shows an adaptively smoothed image
of the entire ACIS-S3 chip ($\sim 346 \times 346$ kpc) region of the cluster.
The image was created with the CIAO algorithm csmooth\footnotemark[2], and
was smoothed to a minimum signal-to-noise of 3 per smoothing beam.
The image was corrected for background and exposure, using the blank sky
background files as discussed in \S~\ref{sec:data}.
There is evidence for structure in the center of the cluster.
A detailed discussion of the
core of the cluster follows below (\S~\ref{sec:core}).

Point sources were detected using the wavdetect algorithm in CIAO.
The significance threshold for detecting a source was set to $10^{-6}$.
This should ensure that there would be $\la$1 spurious detection of a
background fluctuation as a source on the S3 image.
There were 23 sources detected and all of them were confirmed visually.
One of the sources was associated with the center of the cluster.
However, this source was somewhat extended and is likely due, at least in part,
to diffuse emission from the ICM, rather than a central AGN.
Examination of a hard image of this region did not reveal any very
significant point source.

The positions of the X-ray sources were compared with optical/IR positions
from the 2MASS catalog and the USNO-A2.0 catalog.
In the end, the 2MASS catalog gave a smaller dispersion in the positional
offsets than the USNO-A2.0 catalog, and we used the 2MASS positions to
check the X-ray coordinates.
One of the X-ray sources with an optical/IR counterpart was an extended
source near the edge of the chip, and it was not taken into consideration.
As noted above, there was an X-ray source associated with the center of
the cluster. Although it matched the 2MASS position of the central cD galaxy
CGCG 077$-$097 and the position of the core of the radio source
[OL97]1520+087 \citep{owen} associated with the central cD, we did not use this
correspondence as the central source is likely to be due to ICM X-ray
emission rather than an AGN.
There is an X-ray source associated with the cluster galaxy
LEDA 084605, which is also the radio source
[AO95]1520+0849 \citep{ao95}.
Another X-ray source is associated with
the cluster galaxy CGCG 077$-$096, which is also the
radio source [AO95]1520+0845 \citep{ao95}.
The average of the offsets between the positions measured
in the X-ray and the positions from the 2MASS catalog was determined to be
$+0\farcs14$ in RA and $+0\farcs44$ in Dec.
Since the offset in RA was smaller than the dispersion in the offsets,
it was decided that there was no need to shift the observation in RA.
However, the $+0\farcs44$ in Dec offset was applied to the observation
in order to match the X-ray positions to those from the 2MASS catalog.

\section{Radial Surface Brightness Profile} \label{sec:SB}

\begin{figure}[hp]
\vskip 3.3truein
\includegraphics{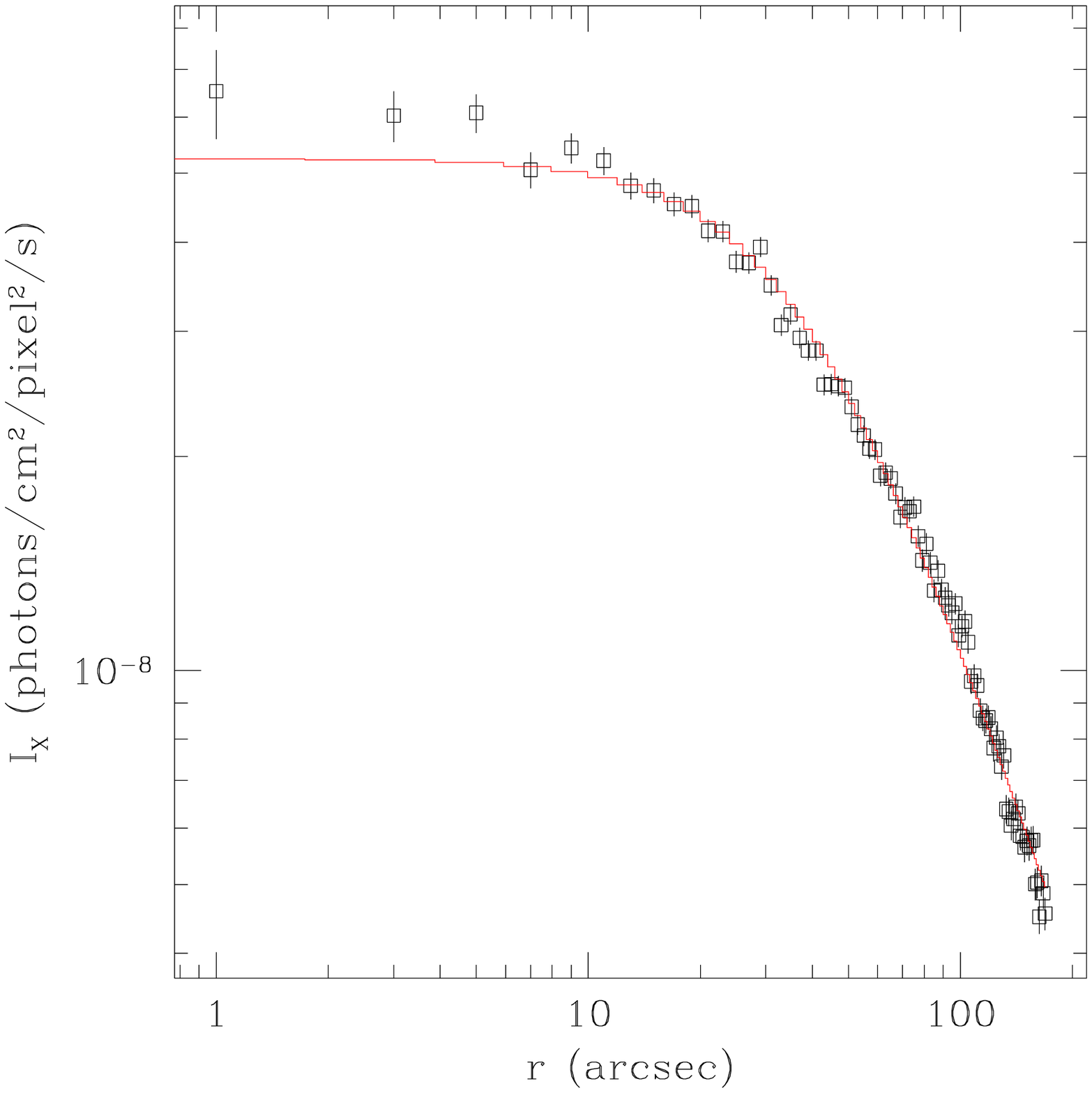}
\vskip 0.5truein
\figcaption{Radial surface brightness profile (out to a radius of
$170\arcsec = 117$ kpc)
and a one-dimensional beta-model fit.
The excess emission seen in the center of the cluster is an indication of
a ``cooling flow.''
\label{fig:SB}}
\end{figure}

The outer isophotes of the cluster are roughly circular, so the surface
brightness should be reasonably represented by a radial profile.
A radial surface brightness profile was obtained for the cluster using the
CIAO tool dmextract. The profile was corrected for exposure and background;
the latter was done using the renormalized blank sky background
(\S~\ref{sec:data}).
Concentric annular regions 2\arcsec\ in radius were used, centered on the
nucleus of the central cD galaxy in the cluster (which coincided with the
peak in the X-ray surface brightness).
The annuli extended out to a radius of 170\arcsec\ (116.62 kpc).
The resulting surface brightness profile is shown in Figure~\ref{fig:SB}.

Two models were used to fit the data in Sherpa\footnotemark[2]:
a one-dimensional beta-model (beta1d) (Figure~\ref{fig:SB}) and
a combination of two one-dimensional beta-models (beta1d + beta1d).
The X-ray surface brightness of a single beta-model is given by
\begin{equation} \label{eq:beta}
I_X ( r )
=
I_0
\left[ 1 + \left( \frac{r}{r_{c}} \right)^2 \right]^{-3 \beta + 1/2}
\, ,
\end{equation}
where $I_0$ is the central surface brightness and $r_c$ is the core
radius.
For the single beta model,
the core radius was determined to be
$r_{c} = 38\farcs9^{+0.2}_{-0.1}$
($26.7^{+0.2}_{-0.1}$ kpc).
The central surface brightness of the fit
was $I_0 = ( 5.22 \pm 0.02 ) \times 10^{-8}$
photons cm$^{-2}$ sec$^{-1}$ arcsec$^{-2}$.
The value of the best-fitting parameter $\beta$ was
$0.430 \pm 0.001$.
The value of $\chi^2$ per degree of freedom for this fit was 2.26.
The fit improves significantly if two beta-models are used (beta1d
+ beta1d).
In this case, the best-fitting parameters for the inner component are:
core radius $r_{1} = 79^{+1}_{-2}$ arcsec ($54 \pm 1$ kpc),
central surface brightness
$I_{1} = ( 2.45 \pm 0.07 ) \times 10^{-8}$ photons cm$^{-2}$ sec$^{-1}$
arcsec$^{-2}$, and
$\beta_{1} = 3.07^{+0.11}_{-0.09}$.
For the outer component, the parameters are:
core radius
$r_{2} = 90\farcs1^{+0.4}_{-0.3}$ ($61.8^{+0.3}_{-0.2}$ kpc),
central surface brightness
$I_{2} = ( 3.12 \pm 0.01 ) \times 10^{-8}$
photons cm$^{-2}$ sec$^{-1}$ arcsec$^{-2}$, and
$\beta_{2} = 0.590 \pm 0.002$.
The value of $\chi^2$ per degree of freedom for this fit was 1.23.

\begin{figure}[hp]
\vskip 2.7truein
\includegraphics{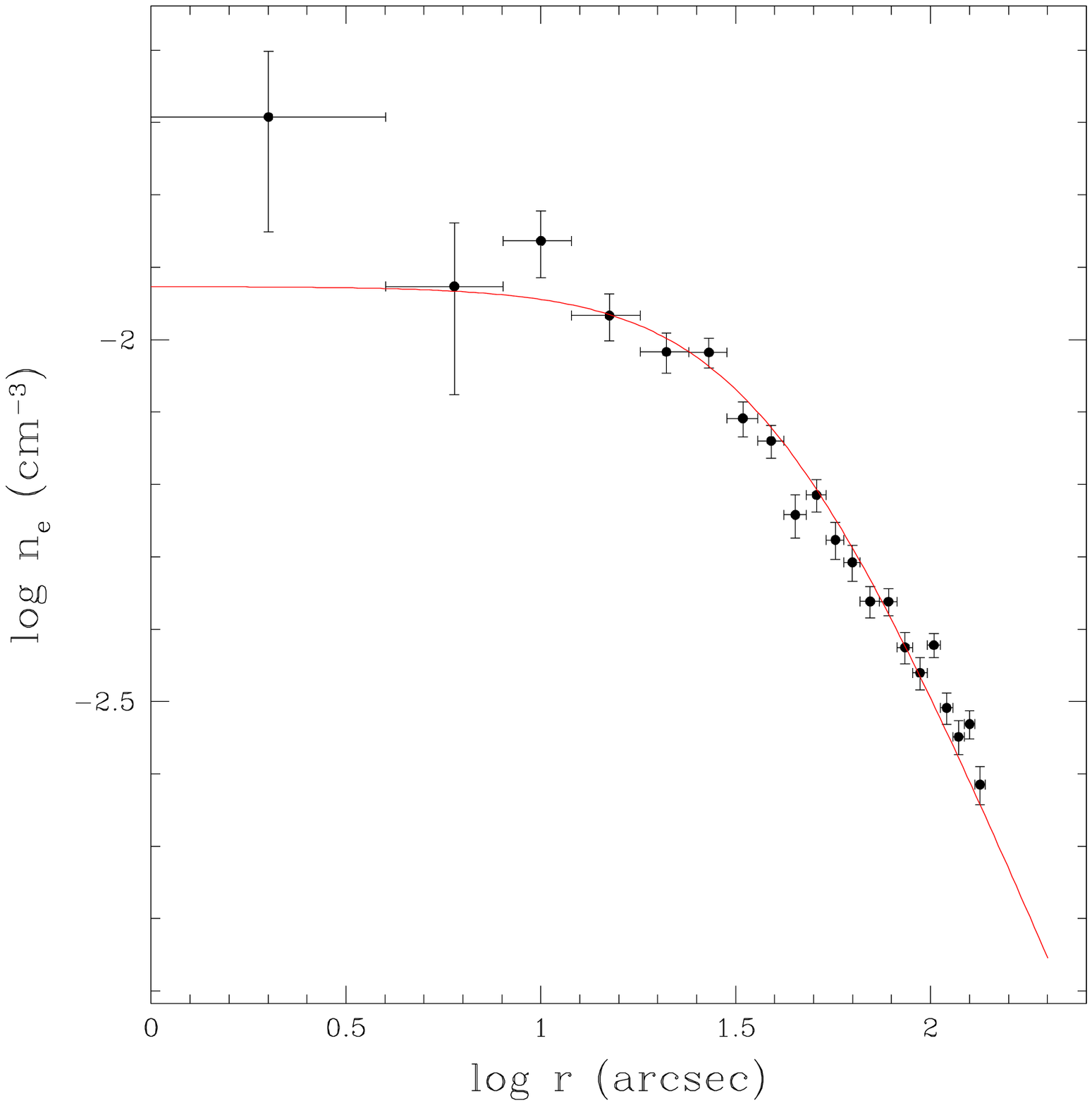}
\vskip 0.6truein
\figcaption{Electron density as a function of radius.
The solid curve represents the values calculated from the single-component
beta-model fit to the surface brightness, while the circles with error bars are
the results of the deprojection.
\label{fig:ne}}
\end{figure}

An estimate of the gas density was derived from the single beta-model
fit to the X-ray surface brightness, and the resulting electron number
density profile is shown in Figure~\ref{fig:ne}.
The X-ray emissivity was converted to an electron density using the
best single temperature fit to the {\it Chandra} total cluster spectrum
(\S~\ref{sec:spectrum}).
The form of the electron density for the beta-model is
\begin{equation} \label{eq:ne}
n_e ( r )
=
n_e ( 0 )
\left[ 1 + \left( \frac{r}{r_{c}} \right)^2 \right]^{-3 \beta / 2}
\, ,
\end{equation}
where the central electron density derived from the best-fit single
component beta-model is
$n_{e} ( 0 ) = 1.18 \times 10^{-2}$ cm$^{-3}$.
The gas density was also determined directly from the X-ray surface
brightness by deprojection.
Again, we used the best single temperature fit to the {\it Chandra} total
cluster spectrum (\S~\ref{sec:spectrum}) to determine the emissivities
and the electron densities.
The deprojection assumed that the X-ray emissivity was constant within
spherical shells whose radii were those of the annuli used to accumulate
the X-ray surface brightness
(Figure~\ref{fig:SB}).
The surface brightness was taken to be zero outside of the observed region,
and this will cause the outermost values of the electron density to be
incorrect.
These values were dropped from the plots.
The resulting deprojected electron density is shown by the points with
error bars in Figure~\ref{fig:ne}.
Note that the deprojection causes the uncertainties to be correlated.
The single component beta model fits the deprojected density reasonably
well except in the center, where the surface brightness is affected by
the cool core and by the structure in the X-ray emitting gas
(\S~\ref{sec:image}).

\begin{figure}[hp]
\vskip 2.7truein
\includegraphics{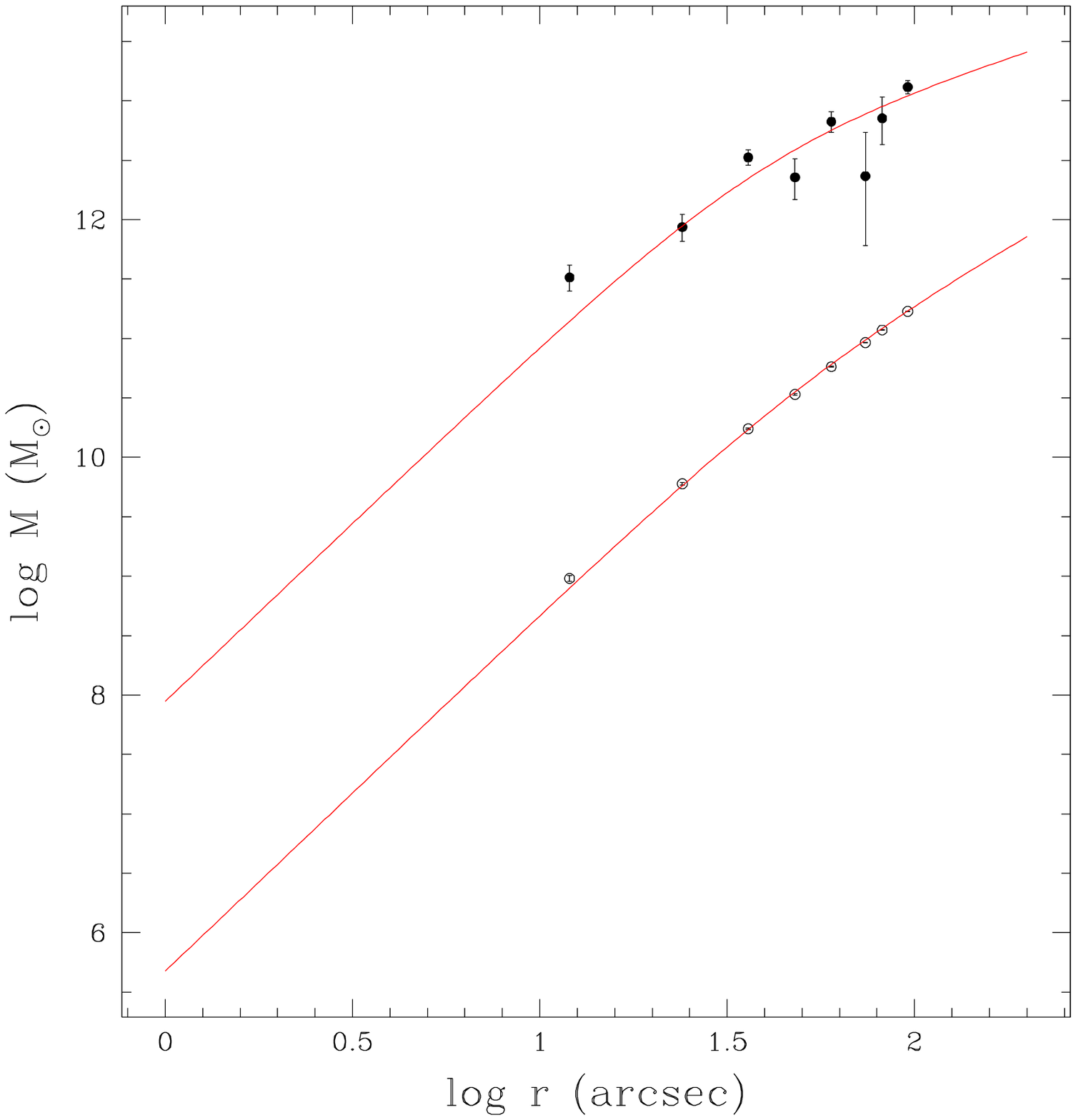}
\vskip 0.6truein
\figcaption{Total cluster mass (upper values) and gas mass (lower values)
as a function of radius.
The circles with error bars are the results from the deprojected gas
density, while the solid curves are calculated from the single-component
beta-model fit to the surface brightness.
\label{fig:masses}}
\end{figure}

These results were used to determine the gas and total mass of the cluster
as a function of the radius.
The gas temperature was assumed to be constant at the value given by the
best single temperature fit to the {\it Chandra} total cluster spectrum
(\S~\ref{sec:spectrum}).
The gas mass is given by:
\begin{equation}
M_{\rm gas}(r) = \int_0^r \rho_{\rm gas}(r) 4 \pi r^2 dr
\, .
\end{equation}
The total mass is given by the condition of hydrostatic equilibrium:
\begin{equation}
M_{\rm tot}(r) =
- \left[ \frac{k T(r) r}{\mu m_H G} \right]
\left(
\frac{d \ln \rho}{d \ln r}
+ \frac{d \ln T}{d \ln r} \right)
\, .
\end{equation}
The errors in the surface brightness profile were determined from the
Poisson errors on the counts in annuli for the observation and for the
blank sky background.
During the deprojection of the surface brightness profile, these errors
were propagated.
Because the deprojection causes the errors on the emissivities in the
different shells to be correlated, the full covariance matrix was
determined. 
These errors, included covariances, were propagated to determine the errors
on the electron density, gas mass, and total mass
in Figures
\ref{fig:ne} \& \ref{fig:masses}.
The error on the total mass included the error on the gas temperature,
which was assumed to be independent of the error in the gas density.

The values of the total and gas masses are shown in Figure~\ref{fig:masses}.
The total mass at $ r = 96\arcsec$ (68.86 kpc) was found to be
$M_{\rm tot} = ( 1.31 \pm 0.06 ) \times 10^{13} \, \msun$.
The value obtained using the parameters from the beta-model fit was
$M_{\rm tot} = 1.12 \times 10^{13} \, \msun$.
The gas mass at the same radius was found to be
$M_{\rm gas} = ( 1.69 \pm 0.03 ) \times 10^{11} \, \msun$, while the
beta-model value was
$M_{\rm gas} = 1.70 \times 10^{11} \, \msun$.
The gas fraction within this radius is thus
$f_{\rm gas} \equiv M_{\rm gas} / M_{\rm tot} = 0.013$.
The low masses and gas fraction are a result of the small radius out to
which these values are determined from the {\it Chandra} S3 data.

\section{Integrated Spectrum} \label{sec:spectrum}

A spectrum was obtained from a circular region, centered on the cluster
center and extending nearly to the nearest edge of the ACIS-S3 chip.
The region was 330\arcsec\ = 226.4 kpc in diameter. The
point sources detected with wavdetect were excluded with the exception
of the source associated with the center of the cluster, since it is
mainly due to structure in the diffuse ICM emission,
rather than an AGN at the center.
As discussed in \S~\ref{sec:data}, we excluded data from the first
observation as the background was higher than normal and included a
different spectral component.
The spectrum, weighted response
functions and background spectrum were obtained using the
CIAO acisspec\footnotemark[2] script.
The data were grouped into bins with a minimum of 25 counts.
The blank-sky observations of M. Markevitch\footnotemark[3]
were used for the background spectrum.
Several fits were made to the spectrum using the
XSPEC Version 11.3.0 \citep{xspec} software package,
including an absorbed single temperature model,
(wabs*apec), an absorbed two-temperature model [wabs*(apec+apec)],
and an absorbed single temperature plus cooling flow model
[wabs*(apec + mkcflow)].
We initially attempted to fit the spectra over the energy range
from 0.3 to 10 keV.
However, all of the fits had significant residuals at low energies.
Allowing the Galactic absorbing column to vary improved the fits slightly,
but these fits had new residuals at higher energies.
We concluded that the residuals were the result of an imperfect modeling
of the absorption due to the build-up of contaminating material
on the optical-blocking filters of the ACIS detectors.
Thus, we restricted the fits to the energy range 0.5--10 keV.
The total number of net counts that were detected in this energy range was
73,808.

The results are given in Table~\ref{tbl:xspec}.
The best fit to the spectrum using a single-temperature model gave an average
temperature of the cluster of $kT = 3.66 \pm 0.09$ keV and an average
abundance of $Z = 0.54^{+0.06}_{-0.05}$ times the solar value.
The single temperature model provided an adequate fit to the spectrum;
the value of $\chi^2$ per degree of freedom was 0.973.
However, the pattern of residuals suggested that a softer component of the
spectrum was missing.

\begin{figure}[hp]
\vskip 3.0truein
\includegraphics{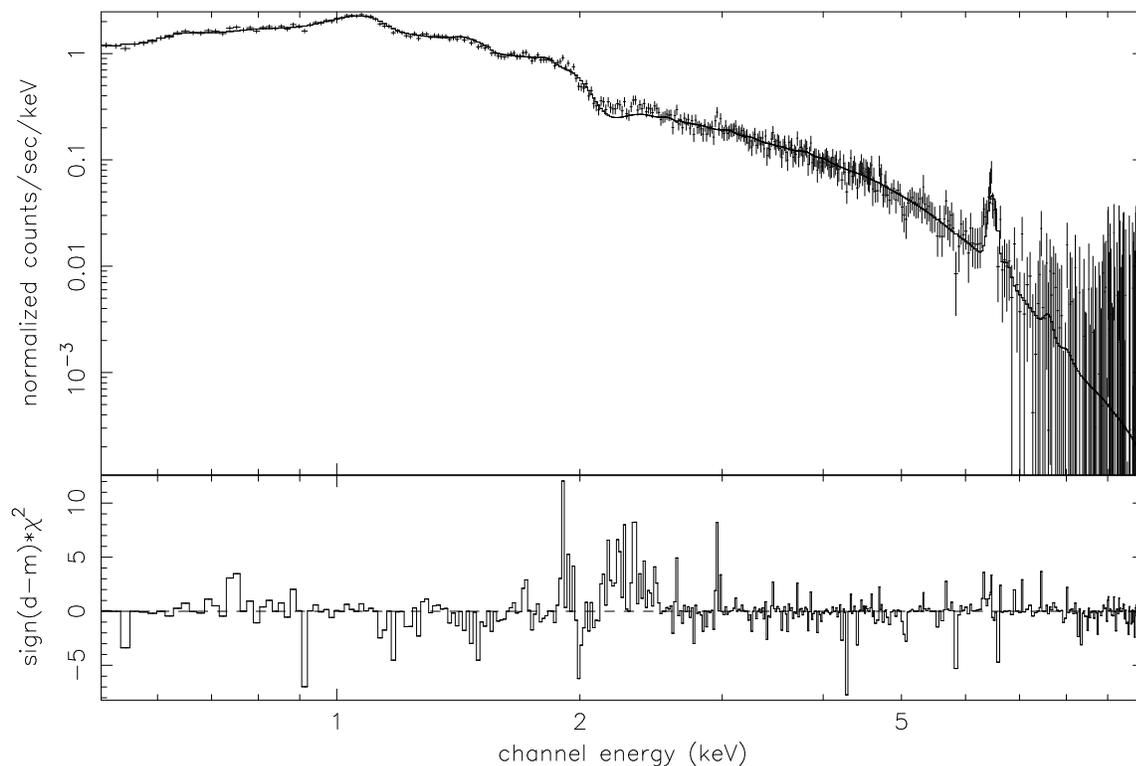}
\vskip 1.0truein
\figcaption{X-ray spectrum of the cluster for a circular region with a radius
of 165\arcsec\ (113.2 kpc)
centered on the radio core in the 0.5 - 10.0 keV energy band.
In the upper panel, the data points are the observed spectrum,
and the solid histogram is the best-fit
two-temperature thermal plasma model with variable absorption
(Table~\ref{tbl:xspec}).
The lower panel gives the residuals to the fits expressed as their
contribution to the values of $\chi^2$, with the sign indicating the
sense of the difference between the data and the model.
The unusually high residuals around 2.0 keV may be due to the telescope
iridium edge.
\label{fig:spec}}
\end{figure}

The fit is improved
with the addition of a second thermal component (apec).
The lower temperature was about a factor of two below the higher
temperature.
The abundances were poorly constrained if the two emission components
were allowed to have different abundances, so we assumed both abundance
values were the same.
The average abundance was $0.50 \pm 0.06$ times the solar value.
The value of $\chi^2$ per degree of freedom was 0.946.
The two-temperature fit to the spectrum is slightly improved if the absorption
is allowed to vary.
The value of the absorption that produces the best fit is slightly higher
than the Galactic value $N_H = ( 3.91 \pm 0.48 ) \times 10^{20}$ cm$^{-2}$;
the Galactic value is $3.03 \times 10^{20}$ cm$^{-2}$.
The spectrum and the fitted model are shown in Figure~\ref{fig:spec}.
The largest residuals are near 2 keV;
it is likely that this is the result of the sharp drop in the telescope
collecting area there due to the absorption edge of iridium.

\begin{deluxetable}{lcccccc}
\tabletypesize{\scriptsize}
\tablewidth{0pt}
\tablecaption{Fits to the Total Spectrum}
\tablehead{
\colhead{Model}& \colhead{$N_H$} & \colhead{$kT_{\rm low}$} & \colhead{$kT_{\rm high}$} & \colhead{Abundance} & \colhead{$\dot M$} & \colhead{$\chi^2$/d.o.f.}
\\
\colhead{} & \colhead{($\times 10^{20}$ cm$^{-2}$)} & \colhead{(keV)} & \colhead{(keV)} & \colhead{(Solar)} & \colhead{($\msun$ yr$^{-1}$)} & \colhead{}
}
\startdata
wabs*apec & (3.03) & \nodata & $3.66^{+0.09}_{-0.09}$ & $0.54^{+0.06}_{-0.05}$ & \nodata & 441/454=0.973 \\
wabs*(apec+apec) & (3.03) & $1.7^{+0.7}_{-0.4}$ & $4.09^{+1.00}_{-0.29}$ & $0.50^{+0.06}_{-0.06}$ & \nodata & 427/452=0.946 \\
wabs*(apec+apec) & $3.9 \pm 0.5$ & $2.0^{+0.5}_{-0.7}$ & $4.24^{+0.55}_{-0.62}$ & $0.44^{+0.06}_{-0.06}$ & \nodata & 418/451=0.927 \\
wabs*(apec+mkcflow) & (3.03) & $1.2^{+0.7}_{-0.6}$ & $4.06^{+0.68}_{-0.28}$ & $0.51^{+0.06}_{-0.06}$ & $20^{+29}_{-16}$ & 430/452=0.951 \\
wabs*(apec+mkcflow) & $4.3 \pm 0.5$ & $0.1^{+1.7}_{-0.1}$ & $3.66^{+0.12}_{-0.14}$ & $0.52^{+0.03}_{-0.05}$ & $4.0^{+1.7}_{-1.9}$ & 420/451=0.932 \\

\enddata
\tablecomments{Values in
parentheses were held fixed in the models.}
\label{tbl:xspec}
\end{deluxetable}

Lastly, we fitted the spectrum with a cooling flow model in a combination
with a single temperature thermal plasma model.
The single temperature model was taken to represent the ambient, uncooled
cluster gas, while the cooling flow model represents gas cooling radiatively
to lower temperatures.
The implied cooling rate was ${\dot{M}} \approx 20$ \msun\ yr$^{-1}$,
but with very large errors which extended almost to zero.
The abundances for the two
models were set to be equal and the higher temperature in the cooling flow
model was set to be equal to the temperature of the thermal model. This was
justified by the assumption that the thermal model represents ambient gas in
the cluster and the cooling flow is the result of this gas cooling in the core.
The f-test indicated that the model with a cooling flow was not
significantly better than the model with a single temperature.
The best-fit value of the lower bound to the cooling flow gas did not
quite include zero temperature, so gas was not cooling to very low
temperature, even in the cooling flow model.
However, when the absorption is
allowed to vary the cooling rate drops significantly and the fit indicates
that there could be gas cooling to very low temperatures.
The absorption in this case is higher than the Galactic value;
this extra absorption may absorb away some of the emission from the gas
cooling to very low temperatures.
Given the large uncertainties in the lower temperature of this model, we
view the cooling rate in this model as providing an upper limit on the
rate of gas cooling to very low temperatures of about 6 \msun\ yr$^{-1}$.
We regard the model with the absorption fixed at the Galactic value as
giving a better measure of the amount of gas cooling from the ambient
cluster temperature down to about 1/3 of the ambient temperature.

\section{Temperature Map} \label{sec:t_map}

\begin{figure}[hp]
\vskip 3.9truein
\includegraphics{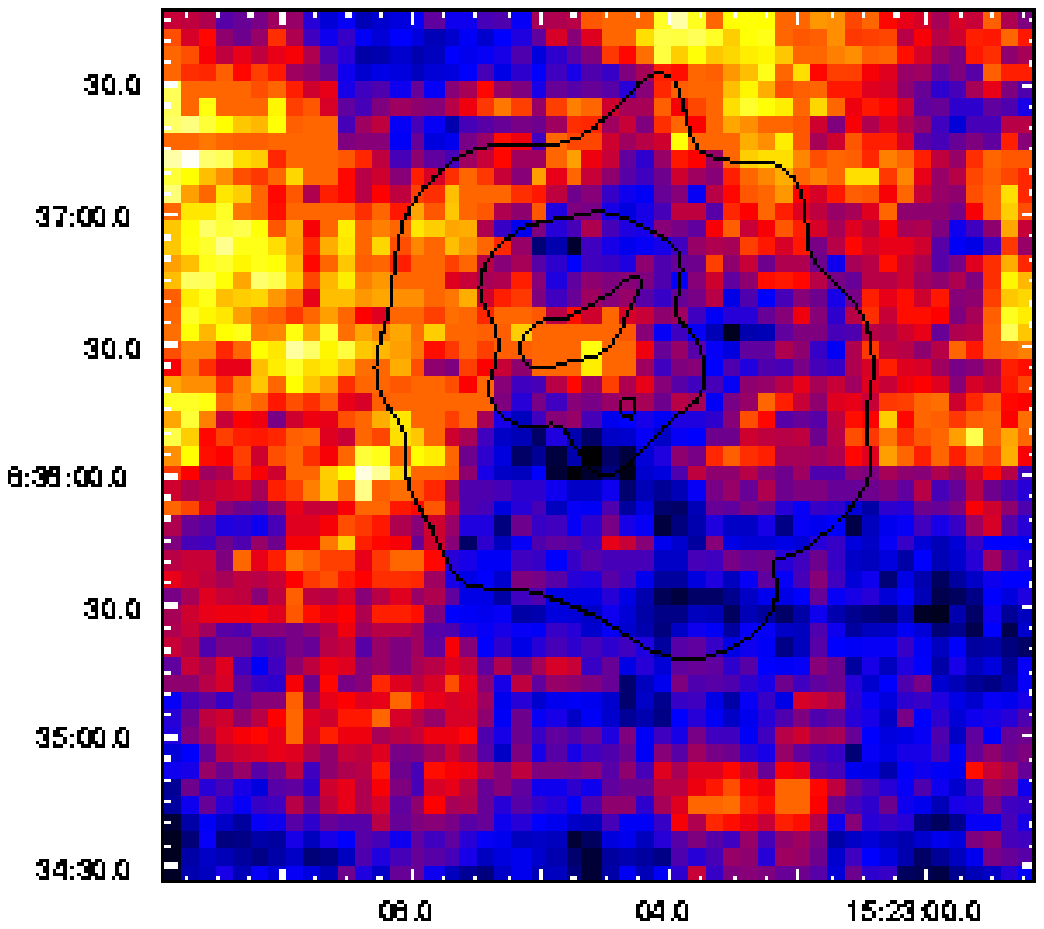}
\vskip 0.2truein
\figcaption{Temperature map of the central $200 \arcsec \times 200 \arcsec
= 137 \times 137$ kpc region of Abell 2063 with X-ray contours from the adaptively 
smoothed image overlaid.
The contour levels are (0.203, 1.16, and 2.12) $\times 10^{-5}$ 
photons s$^{-1}$ cm$^{-2}$ pixel$^{-2}$.
The dark blue (near black) colors show cool gas (2.63 keV)
and the light yellow (near white) colors are hot gas (4.34 keV).
The gas surrounding the radio lobe north of the center appears to be hot,
which may be evidence of a shock driven into the gas by the radio lobe.
\label{image:t_map}}
\end{figure}

A temperature map of the center of the cluster is shown in
Figure~\ref{image:t_map}. It was generated by fitting individual spectra
with
ISIS\footnote{http://space.mit.edu/ASC/ISIS/}
\citep{houck}
in the 0.5-10.0 keV energy range, with the condition that there
are at least 1000 counts including background in each spectrum. Point
sources were excluded before the map was created. A thermal plasma model
(wabs*apec) was used for the fit to the spectra and each spectrum was
binned to a minimum of 25 counts per bin.
The absorption was fixed at the Galactic value.
The initial parameters used were
taken from the best single temperature fit to the cluster-integrated
spectrum. The map is an array of $50 \times 50$ boxes, where each box
represents the center of a region (that contains at least 1000 counts)
from which a spectrum was extracted and fitted.

The temperatures in the map range from 2.63 to 4.34 keV.
Errors on the temperature values are typically 10--15\%.
The coolest gas is found surrounding the center of the cluster, which is
what is typically seen in cool core clusters.
However, there is also hot gas near the center which is associated with
X-ray structures in the central region.
These features are discussed in more detail below (\S~\ref{sec:core}).

\section{X-ray/Radio Interactions in the Cluster Core} \label{sec:core}

\begin{figure}[hp]
\vskip 4.1truein
\includegraphics{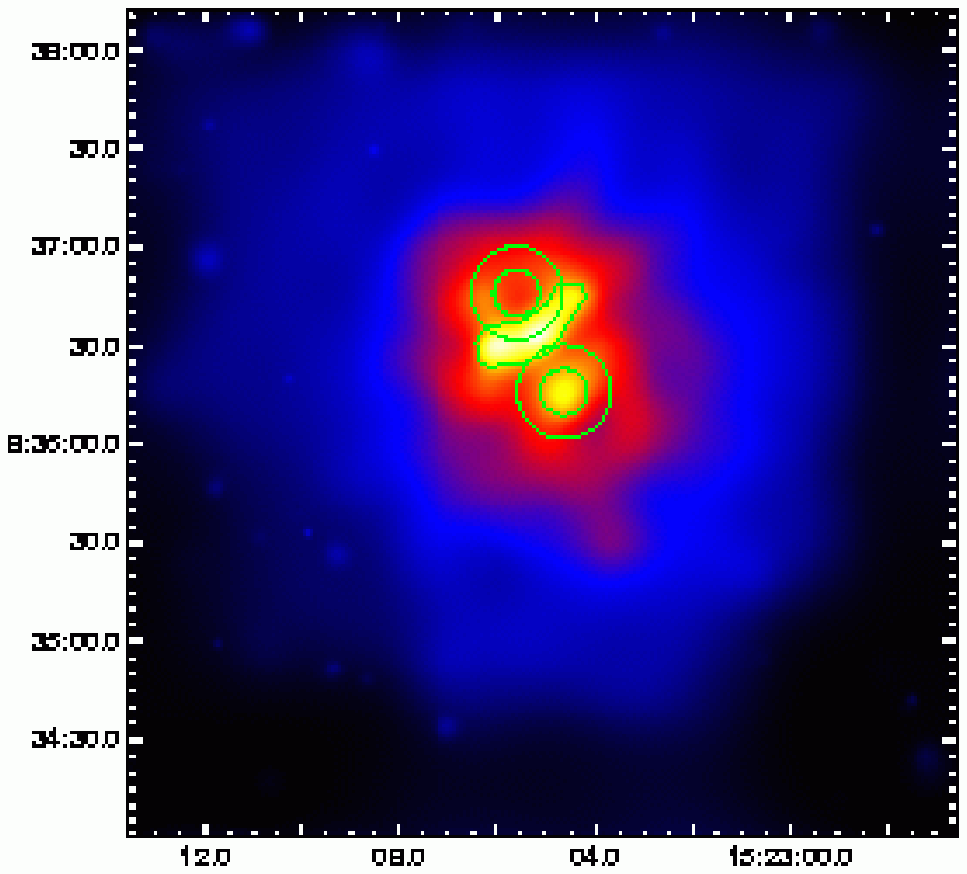}
\vskip 0.1truein
\figcaption{Adaptively smoothed X-ray image of the central $\sim 4\farcm2
\times 4\farcm2$ region of Abell 2063 in the $0.3-10.0$ keV energy band.
The image has been corrected for background and exposure.
The image shows a central ``bar'' of enhanced emission centered on the
nucleus of the central cD galaxy.
There is a ``hole'' of reduced X-ray emission north of the bar,
which is surrounded by a possible ``shell'' of enhanced emission
(to the east, west, and south, but perhaps not to the north);
the southern portion of the shell merges with the bar.
There also is a bright ``blob'' of emission
to the south of the center, and several other weaker features are
evident. The regions that were defined for these features and from
which spectra was extracted (results are given in Table~\ref{tbl:regions}) 
are shown in green.
\label{image:smoothed}}
\end{figure}

Figure~\ref{image:smoothed} shows an adaptively smoothed image
of the central $252\arcsec \times 252\arcsec = 173 \times 173$ kpc of the
cluster.
The image was created in a similar manner as the adaptively smoothed image
of the entire ACIS-S3 chip (Figure~\ref{image:smoothed_image_S3}) as
discussed in \S~\ref{sec:image}. It
was smoothed to a minimum signal-to-noise of 3 per smoothing beam.
The image was corrected for background and exposure, using the blank sky
background files as discussed in \S~\ref{sec:data}.

The core of the cluster exhibits a complex structure.
There is a bright knot of emission associated
with the nucleus of the central cD galaxy.
There are bright regions to the east and northwest of this nucleus which
may form a central ``bar''.
There is also a bright ``blob'' of emission to the south of the nucleus.
There are also several regions of surface brightness deficit,
including a ``hole'' in the emission to the north of the center of the
cluster.
There are bright regions of emission to the east and west of the hole,
which, together with the central bar, may form a ``shell'' surrounding the
northern hole.
There is a possible second hole to the south of the nucleus beyond the
southern blob.

We determined the X-ray count rates and surface brightnesses for the northern
hole and shell to assess their statistical significance.
The hole was taken to be a circular region with a diameter of 14\farcs2 =
9.7 kpc.
The full shell was represented as a circular annulus with inner and outer
radii of $7\farcs1 = 4.88$ kpc and $14\farcs2 = 9.7$ kpc, respectively.
We also consider a partial shell, consisting of the same annulus but without
the northern portion.
This included all of the regions which appear to be bright in
Figure~\ref{image:smoothed}.
The integrated average surface brightness of this incomplete shell
was $0.78 \pm 0.03$ counts pixel$^{-2}$, while the corresponding value
for the hole was $0.67 \pm 0.03$ counts pixel$^{-2}$.
The difference is significant at the 3$\sigma$ level.
If the entire circular annulus shell is included, the difference is
slightly less significant.
Thus, there is evidence that the shell is real, but it may not be complete
on the top end.
Similarly, the southern X-ray blob was compared to the region surrounding
it, and was found to be brighter at the 2.5 $\sigma$ level.

\begin{figure}[hp]
\vskip 4.0truein
\includegraphics{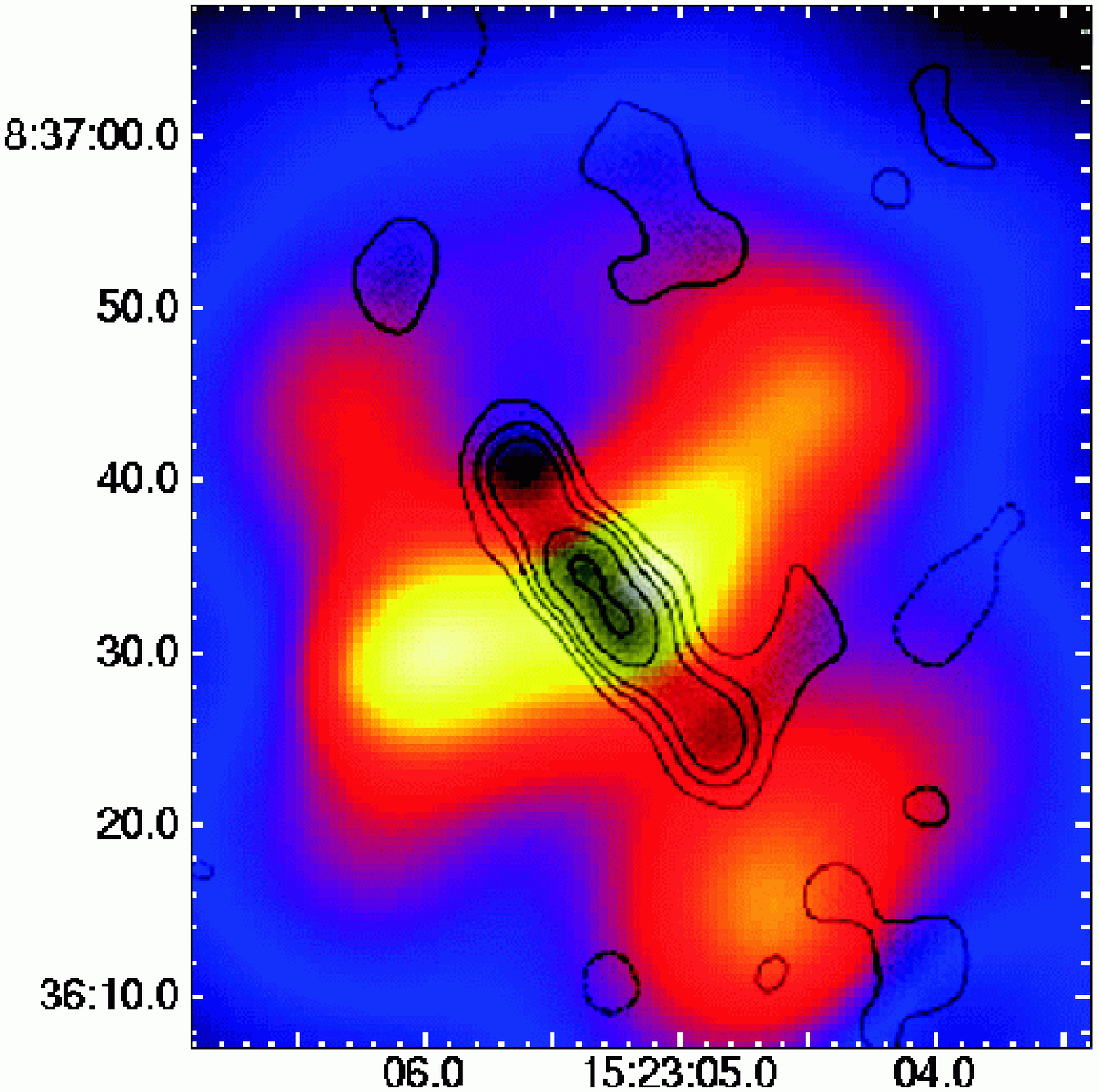}
\figcaption{Central $51\farcs6 \times 60\arcsec = 35.4 \times 41.2$ kpc
region of
the adaptively smoothed $Chandra$ image shown in Figure~\ref{image:smoothed}.
The contours show the 1400 MHz radio
emission of [OL97]1520+087 from \citet{owen}. The X-ray emission shows
the elongated central bar, and the bright blob of emission to the south.
The bright radio core is co-incident with the brightest region of the
X-ray core. The radio jets propagate perpendicular to the central bar,
and extend out to roughly 12\arcsec\ ($\sim$8.2 kpc).
The jet that extends to the north-east partially fills the northern X-ray hole.
\label{image:radio}}
\end{figure}

Figure~\ref{image:radio} shows the contours of the 1.4 GHz radio image of
the central cD (radio source [OL97]1520+087) from \citet{owen}
overlaid on the center of the {\it Chandra} adaptively smoothed image.
The radio core coincides with the nucleus of the central cD galaxy and
the brightest knot of X-ray emission at the center of the cluster.
The radio source is extended northeast-to-southwest, which is perpendicular
to the central X-ray bar. The northern X-ray hole is partially filled by radio
emission. To the south, the radio source also is located in a region of reduced
X-ray emission, but with the southern X-ray blob lying just beyond the end of
the radio source in this image.

It is interesting to compare the structures in the X-ray and radio images
with the temperature map of the center of the cluster
(Fig.~\ref{image:t_map}).
The bright central bar appears to be hotter than the surrounding gas.
The southern blob is of intermediate temperature.
The northern X-ray hole is cooler than average, and the shell surrounding
the hole is hotter.
The temperature structure appears quite different from that seen in most
other clusters with ``radio bubbles''
\citep[e.g.,][]{blanton03}.
In most cases, the center of the cluster is cool, and the shells surrounding
the radio lobes are cool.
Generally, the X-ray holes appear hot, but consistent with the
temperature of the outer cluster gas projected onto the holes.
The anticorrelation between the X-ray surface brightness and temperature seen
in most cool core clusters is qualitatively consistent with near pressure
equilibrium.
On the other hand, the positive correlation between temperature and surface
brightness in Abell~2063 suggests that there are large pressure variations,
and may indicate that the center of this cluster is more dynamically
active. This may be an evidence of
the ICM being shocked by the radio source as is predicted by some theoretical
models
\citep[e.g.,][]{heinz98}.
Recently, a number of cases of apparent shocks which surround radio bubbles have
been observed with {\it Chandra}
\citep{mcnamara05,nulsen05a,nulsen05b}.

In order to accurately quantify any differences in the temperature of these
regions near the center of Abell~2063, we extracted their spectra and fitted
them to absorbed single temperature models; the results are shown in
Table~\ref{tbl:regions}.
In all cases, the absorbing column was set to the Galactic value.
(Allowing it to vary did not change the results significantly.)
The spectra were grouped to a minimum of 25 counts per bin,
and were fitted with XSPEC.
The blank sky background was used (\S~\ref{sec:data}).
The spectra were fit in the photon energy range of 0.5--10 keV.

First, spectra were
extracted from regions corresponding to the northern X-ray hole and the
shell surrounding it.
The best fit to the spectrum of the northern hole gave a value for the
temperature of the gas in that region of $kT = 2.7^{+1.3}_{-0.5}$ keV and
an abundance of $Z = 0.7^{+2.0}_{-0.4}$ times the solar value.
However, these values were poorly determined since there were only 405 net
counts in the spectrum.
Another fit was made to this spectrum with the abundance set to
the value from the single temperature fit to the total X-ray spectrum,
which was 0.54 times the solar value.
The value for the temperature obtained with this fit was
$kT = 2.6^{+0.8}_{-0.5}$ keV.
The best fit to the spectrum of the northern shell gave a value for
the temperature of the gas in that region of
$kT = 3.6^{+0.5}_{-0.4}$ keV and an
abundance of $Z = 0.9^{+0.8}_{-0.4}$ times the solar value. Setting the
abundance to the global value determined from the single temperature fit to the
total X-ray spectrum we derive a temperature of the gas in that region of
$kT = 3.5^{+0.5}_{-0.4}$ keV.
There were a total of 1309 net counts detected in this region.
This agrees with the result from the temperature map that the shell is
hotter than the hole, but the difference is not significant at the 90\%
confidence limit.
The higher temperature in the shell may indicate
that we are observing one of the rare cases in which the ICM is heated by
the radio source through shocks into the X-ray gas.
This may also imply that the radio source is a relatively young source.
There are observational indications of shock heating by radio sources
in other clusters as well
\citep{mcnamara05,nulsen05a,nulsen05b}.

The spectrum was also extracted from the southern blob of X-ray emission,
and from an annular region surrounding it.
The best fit spectrum of the southern blob gave a temperature of
$kT = 3.9^{+1.2}_{-0.7}$ keV and an abundance of
$Z = 0.9^{+2.3}_{-0.7}$ times the solar value.
However, these values were poorly determined since there were only 473 net
counts in the spectrum. For this reason the abundance was fixed to the global
value as discussed above, which gave a temperature for the blob of
$kT = 3.8^{+1.2}_{-0.8}$ keV.
The surrounding region had a temperature of
$kT = 3.8^{+0.6}_{-0.5}$ keV and an abundance of $Z = 1.6^{+1.7}_{-0.6}$
times solar.
There were a total of 1296 net counts in this spectrum. When the abundance was
fixed to the global value the derived temperature was
$kT = 3.4^{+0.8}_{-0.4}$ keV

The spectrum from the central bar in the cluster was also extracted,
and gave a best-fit temperature of $kT = 3.7^{+0.5}_{-0.4}$ keV and an
abundance of $Z = 0.9^{+0.7}_{-0.4}$ times the solar value.
There were a total of 1528 counts detected in this region and in the
energy range used for the fit.

\begin{deluxetable}{lcccccc}
\tabletypesize{\scriptsize}
\tablewidth{0pt}
\tablecaption{Fits to the Spectra of Individual Regions}
\tablehead{
\colhead{Region} & \colhead{$N_H$} & \colhead{$kT$} & \colhead{Abundance} & \colhead{$\chi^2$/d.o.f.}
\\
\colhead{} & \colhead{($\times 10^{20}$ cm$^{-2}$)} & \colhead{(keV)} & \colhead{(Solar)} & \colhead{}
}
\startdata
northern hole & (3.03) & $2.6^{+1.3}_{-0.5}$ & $0.7^{+2.0}_{-0.4}$ & 11.0/12=0.916 \\
northern hole & (3.03) & $2.5^{+0.8}_{-0.5}$ & (0.54) & 11.5/13=0.886 \\
northern shell & (3.03) & $3.6^{+0.5}_{-0.4}$ & $0.9^{+0.8}_{-0.4}$ & 38.8/41=0.947 \\
northern shell & (3.03) & $3.5^{+0.5}_{-0.4}$ & (0.54) & 41.2/42=0.981 \\
southern blob & (3.03) & $3.9^{+1.2}_{-0.7}$ & $0.9^{+2.3}_{-0.7}$ & 14.0/15=0.931 \\
southern blob & (3.03) & $3.8^{+1.3}_{-0.8}$ & (0.54) & 14.6/16=0.914 \\
southern shell & (3.03) & $3.8^{+0.6}_{-0.5}$ & $1.6^{+1.7}_{-0.6}$ & 50.1/40=1.252 \\
southern shell & (3.03) & $3.4^{+0.8}_{-0.4}$ & (0.54) & 59.4/41=1.448 \\
central bar & (3.03) & $3.7^{+0.5}_{-0.4}$ & $0.9^{+0.7}_{-0.4}$ & 56.9/47=1.210 \\
central bar & (3.03) & $3.7^{+0.5}_{-0.5}$ & (0.54) & 59.4/48=1.237 \\
\enddata
\tablecomments{Values in
parentheses were held fixed in the models.}
\label{tbl:regions}
\end{deluxetable}

\section{Discussion} \label{sec:disc}

Figure~\ref{image:radio} suggests that Abell~2063 is another example of
a cool core cluster with ``radio bubbles'':
a hole in the X-ray gas occupied by radio plasma, surrounded by a shell
of compressed X-ray gas which is the material displaced from the hole.
Since the shell/hole structure is much clearer for the northern
radio lobe, we mainly discuss this feature.
The nature of the southern X-ray blob is unclear;
it might indicate that the southern radio jet is either entraining or
compressing denser X-ray gas, or this blob may be a particularly dense
region on the front or back surface of a southern radio bubble.

In order to assess the energetics of the radio source, we begin by
determining the pressure in the northern shell.
Since the normalization of the thermal plasma models fitted to the
spectrum of each of the individual regions is proportional to the square
of the electron density, we can determine the density, temperature, and
pressure from the spectral fits.
The best-fit model to the spectrum of the northern shell
gave a temperature of $kT = 3.52^{+0.54}_{-0.42}$ keV, and a normalization of
$K= ( 2.62 \pm 0.18 ) \times 10^{-4}$ cm$^{-5}$.
The normalization $K$ is given by
\begin{equation}
K=\frac{10^{-14}}{4\pi D_A^2 (1+z)^2} \int n_H n_e dV \, .
\end{equation}
Here, $D_A$ is the angular diameter distance to the cluster,
$z$ is the redshift, and
$n_e$ and $n_H$ are the electron and hydrogen densities in cm$^{-3}$,
respectively.
We assume that the emission is from a spherical shell with inner radius 
of $7\farcs1 = 4.88$ kpc and outer radius of $14\farcs2 = 9.7$ kpc.
We find an electron density of $n_e = 0.033\pm 0.002$ cm$^{-3}$ and a
pressure of
$P = ( 3.6 \pm 0.6 ) \times 10^{-10}$ dyn cm$^{-2}$.

This radio source has total 1.4 GHz flux of 13 mJy, which corresponds to a
power of $P_{1.4} = 3.6 \times 10^{22}$ W Hz$^{-1}$
\citep[converted to our cosmology]{owen}.
Unfortunately, we do not have any information on the flux for the northern
lobe alone, so we will determine average properties for the entire radio
source.
We assume that the source is a cylinder with a height of 23\arcsec\ and
a radius of 6\arcsec.
The radio spectral index is not known on the scale of the X-ray bubble, so
we assume a value of $\alpha = -1.0$.
We assume the radio plasma has a volume filling factor of unity,
and that the ratio of the energy in ions to electrons is also unity.
Using the observed flux density of 13 mJy and frequency of 1.4 GHz, we find
an average minimum energy radio luminosity of $L_{\rm radio} = 5.83 \times 10^{40}$ 
ergs s$^{-1}$ and
an average minimum energy nonthermal pressure of
$P_{\rm min} = 4.45 \times 10^{-12}$ dynes cm$^{-2}$ for the radio source.
This pressure is about 80 times lower than the pressure in the surrounding 
shell determined from the X-ray data.
\citet{odea} give minimum pressure for the radio source
$P_{\rm min} = 38 \times 10^{-12}$ dynes cm$^{-2}$,
based on their 1.4 GHz radio map and converted to our cosmology.
This probably applies only to the central components of the radio
source in Figure~\ref{image:radio}, and the pressure is probably lower
in the northern radio bubble.
In any case, this value is still an order of magnitude lower than the
X-ray derived pressure for the shell.
Assuming the radio source did displace the X-ray gas and create the
X-ray hole, the total pressure in the radio lobe must be at least as
large as that in the shell.
Thus, this suggests that the radio lobe has some additional source of
pressure support.
Disagreements in the values
derived for the X-ray and radio pressures have been found in other clusters
as well,
such as Abell 2052 \citep{blanton03} and Abell 262 \citep{blanton04}.

The total energy in the radio lobe can be determined from the energy
needed to evacuate the gas from the hole, plus the internal energy in the
hole \citep[e.g.,][]{churazov02}.
This is given by
\begin{equation}
E_{\rm rad} = \frac{1}{\gamma - 1} P V + P V = \frac{\gamma}{\gamma - 1} P V,
\end{equation}
where $V$ is the volume of the hole and
$\gamma$ is the adiabatic index of the material filling the hole.
Specifically,
$\gamma = \frac{5}{3}$ for
nonrelativistic gas and $\gamma = \frac{4}{3}$ for relativistic gas.
The radius of the northern hole is $7\farcs1 = 4.9$ kpc.
We assume the total pressure in the hole equals that of the shell,
$P = 3.6 \times 10^{-10}$ dyn cm$^{-2}$
as derived above.
This gives energy of
$E_{\rm rad} = 1.29 \times 10^{58}$
ergs for nonrelativistic gas and
$E_{\rm rad} = 2.06 \times 10^{58}$
ergs for relativistic gas.
Under the assumption that the southwestern radio lobe supplies the same
amount of energy, we can compute the total energy
output of the radio source required to inflate the two cavities as
$E_{\rm rad} = 2.58 \times 10^{58}$
ergs for nonrelativistic gas and
$E_{\rm rad} = 4.12 \times 10^{58}$
ergs for relativistic gas.
Radio sources are known to be episodic in nature,
with a typical timescale of
$\simeq 10^{8}$ yr.
If we assume that the radio source in Abell 2063 is similar in nature,
we can estimate the average total power output from the source to be
$8.17 \times 10^{42}$ ergs s$^{-1}$
for nonrelativistic gas and
$1.31 \times 10^{43}$ ergs s$^{-1}$
for relativistic gas.
This mechanical power, required of the AGN
to inflate the two holes, is $\sim 140$ to $\sim 225$ times larger
than the radio luminosity determined above.
This result is consistent with the finding of \citet{birzan04} [eq.~(6) 
in their paper].

The rate of radiative energy loss from the cooling flow can be estimated
from the cooling flow spectral fits.
The luminosity of the cooling gas is given by
\begin{equation}
L_{\rm cool} = \frac{5}{2} \frac{k T}{\mu m_{p}} \dot M \, ,
\end{equation}
where $k T$ is the temperature of the ICM outside of the cooling region,
$\mu$ is the mean molecular mass, and $\dot M$ is the cooling rate.
This assumes the gas cools isobarically to low temperatures.
The energy radiated in cooling down to $T_{\rm low}$ can be estimated
by subtracting the emission at lower temperatures.
Using the best-fit parameters $k T_{\rm high} - k T_{\rm low} = 2.9$ keV and
$\dot M = 20 \, \msun$ yr$^{-1}$ from the cooling flow model, we find
$L_{\rm cool} = 1.5 \times 10^{43}$ ergs s$^{-1}$.
This corresponds to the luminosity of gas cooling
down from 4.06 to 1.17 keV.
As we can see this value is very similar to the
power output from the radio source, which implies that the energy produced
by the radio source is sufficient to offset the cooling flow.
A similar result has been found for many other clusters with radio bubbles
\citep[e.g.,][]{david01,blanton03,birzan04}.

Using the parameters from the best-fitting single
temperature model of the spectrum of the northern shell,
we determined the isobaric cooling time of the shell.
This was done within XSPEC using the same apec emission model which was
used to model the spectrum.
This gave $t_{\rm cool} = 9 \times 10^{8}$ yr.
This is longer than the assumed repetition timescale of the radio source,
which suggests that the radio source is powerful enough to
balance cooling for an even longer period, if the source is truly episodic.
On the other hand, the synchrotron lifetime of the radio
source was estimated to be $2 \times 10^7$ yr, which is shorter than the
cooling time.
Thus, it is likely that the radio plasma within the northern radio lobe has
a very steep spectral index, and that the bubble is maintained by
another pressure source beyond the radio emitting electrons.
Of course, the discrepancy between the minimum energy radio pressure and
the X-ray pressure in the shell also pointed to this.
Low frequency radio observations of this system would be useful to search
for more extended radio emission from within the northern radio bubble, and
possibly in the south as well.

\section{Conclusion} \label{sec:concl}

The {\it Chandra} observation and analysis of the core of the cooling flow
cluster Abell 2063 has been presented above. The X-ray data reveals
complex structure near the center of the cluster, which consists of a
central bar of high surface brightness, a blob of emission to the south
of the center of the cluster, and a depression in the emission to the
north of the center of the cluster which is surrounded by a shell of
higher surface brightness. The position of the hole in the X-ray emission
is coincident with the northeastern radio lobe of the radio source
associated with the central cD galaxy. The temperature map of the
central region of the cluster and detailed spectral analysis of these
features suggest that the northern shell of emission is hotter than
the surrounding gas. This may be evidence that a shock has been
driven into the gas by the radio source. The knot of emission to the
south appeared hotter as well, but it is surrounded by a cooler region.

 From the surface brightness profile of the cluster, we determined the
total mass and the gas mass out to radius $r = 66$ kpc.
We found $M_{\rm tot} = ( 1.31 \pm 0.06 ) \times 10^{13} \, \msun$ and
$M_{\rm gas} = ( 1.69 \pm 0.03 ) \times 10^{11} \, \msun$.

A cooling flow model fitted to the spectrum of the cluster revealed that gas
near the center of the cluster is cooling from 4.06 keV down to 1.17 keV at
a mass deposition rate of
$\dot M = 20^{+29}_{-16} \, \msun$ yr$^{-1}$.
The abundance was found to be 0.51 times the solar value. There was no strong
indication of gas cooling to very low temperatures.

The estimated power output of the radio source in the center of Abell
2063 was found to be sufficient to offset the cooling flow as is the case
with sources in other clusters. The minimum energy nonthermal pressure
provided by the radio source was found to be orders of magnitude lower and
insufficient to balance the pressure exerted by the gas from the bright
shell surrounding the hole. There is some evidence that the radio source
is expanding rapidly and driving a shock into the X-ray gas.
This suggests that the radio source is relatively young, and has
just started its current period of activity.
Low frequency radio observations would be useful to search for more
extended radio emission within the northern radio bubble, and for ghost
bubbles due to previous episodes of AGN activity.

\acknowledgments
We thank Greg Sivakoff who provided us with his data preparation and reduction
software and Marios Chatzikos who helped us with the creation of the 
temperature map.
Support for this work was provided by the National Aeronautics and
Space Administration primarily through {\it Chandra} award
GO5-6126X,
but also through GO4-5133X, GO4-5137, and GO5-6081X.
Some support also came from NASA XMM-Newton award NNG04GO80G.

\end{document}